\documentclass[journal=apchd5,manuscript=article,layout=twocolumn]{achemso}

\usepackage[version=3]{mhchem} 
\usepackage{physics}
\usepackage{amsmath,amssymb}
\usepackage{todonotes}
\usepackage{soul}
\usepackage[normalem]{ulem}
\usepackage[T1]{fontenc}
\usepackage[utf8]{inputenc}

\usepackage{soul}
\usepackage{hyperref}
\hypersetup{
    colorlinks=true,
    linkcolor=blue,
    filecolor=blue,
    urlcolor=blue,
   citecolor=blue,
}
\newcommand{\ao}[1]{\textcolor{cyan}{#1}} 

\DeclareMathOperator{\Lop}{\hat{L}}

\author{Wouter Verstraelen}
\affiliation{Majulab, International Joint Research Unit UMI 3654, CNRS, Université Côte d’Azur, Sorbonne Université, National University of Singapore, Nanyang Technological University, Singapore, Singapore 117543}
\alsoaffiliation{Division of Physics and Applied Physics, School of Physical and Mathematical Sciences, Nanyang Technological University, Singapore, Singapore}
\email{verstraelenwouter@gmail.com}
\author{Stanis\l aw \'Swierczewski}
\affiliation{Institute of Experimental Physics, Faculty of Physics,\\University of Warsaw, ul. Pasteura 5, PL-02-093 Warsaw, Poland}
\alsoaffiliation{Institute of Physics, Polish Academy of Sciences,\\ Aleja Lotnik\'ow 32/46, PL-02-668 Warsaw, Poland}
\author{Andrzej Opala}
\affiliation{Institute of Experimental Physics, Faculty of Physics,\\University of Warsaw, ul. Pasteura 5, PL-02-093 Warsaw, Poland}
\alsoaffiliation{Institute of Physics, Polish Academy of Sciences,\\ Aleja Lotnik\'ow 32/46, PL-02-668 Warsaw, Poland}

\author{Andrew Haky}
\author{Matteo Gadani}
\affiliation{Laboratoire Kastler Brossel, Sorbonne Université, CNRS, ENS-Université PSL, Collège de France, 4 Place Jussieu, Paris, 75005, France}
\author{Huawen Xu}
\altaffiliation{Beijing Academy of Quantum Information Sciences, Beijing, 100193, P.R. China}
\affiliation{Division of Physics and Applied Physics, School of Physical and Mathematical Sciences, Nanyang Technological University, Singapore, Singapore}

\author{Oleksandr Kyriienko}
\affiliation{School of Mathematical and Physical Sciences, University of Sheffield, Sheffield S10 2TN, United Kingdom}
\alsoaffiliation{Department of Physics and Astronomy, University of Exeter, Stocker Road, Exeter EX4 4QL, United Kingdom}
\author{Micha{\l} Matuszewski}
\affiliation{Center for Theoretical Physics, Polish Academy of Sciences,\\  Aleja Lotnik\'w 32/46, PL-02-668 Warsaw, Poland}
\alsoaffiliation{Institute of Physics, Polish Academy of Sciences,\\ Aleja Lotnik\'ow 32/46, PL-02-668 Warsaw, Poland}
\author{Alberto Bramati}
\affiliation{Laboratoire Kastler Brossel, Sorbonne Université, CNRS, ENS-Université PSL, Collège de France, 4 Place Jussieu, Paris, 75005, France}
\author{Timothy C.H. Liew}
\affiliation{Division of Physics and Applied Physics, School of Physical and Mathematical Sciences, Nanyang Technological University, Singapore, Singapore}
\alsoaffiliation{Majulab, International Joint Research Unit UMI 3654, CNRS, Université Côte d’Azur, Sorbonne Université, National University of Singapore, Nanyang Technological University, Singapore, Singapore 117543}
\email{tchliew@gmail.com}

\title{Spectroscopy on a single nonlinear mode recognizes quantum states}

\abbreviations{IR,NMR,UV}

\begin{document}








\newpage{}
\clearpage{}
\begin{abstract}
Characterising optical quantum states is essential for the development of quantum technologies. While traditional approaches to perform full quantum state tomography are often experimentally demanding, neuromorphic architectures may provide an effective alternative. In this work, we demonstrate how a quantum  nonlinear driven-dissipative mode is sufficient to act as a quantum reservoir. By analyzing the occupations at different frequencies in the emission spectrum, a linear regression suffices in many cases to recognize the relevant parameters of incident squeezed states. Beyond highlighting the general potential of this approach under continuous driving, we illustrate its effectiveness in an explicit nontrivial example where the source is a degenerate optical parametric oscillator (OPO), coupled to a nonlinear polariton microcavity. 
\end{abstract}

\section{Introduction}


Neural networks play an increasingly important role in modern electronics, where dedicated neuromorphic chips seek to complement standard von Neumann devices and enhance AI technologies \cite{Kudithipudi2025,Schuman2022}. In particular, neural networks excel at extracting key features and recognizing patterns in large datasets, while remaining robust to errors and missing data.
The objective is not to replace existing machines, but rather complement them. The rise of neural networks also offers an opportunity to revisit other hardware physical platforms including photonics \cite{Momeni2025,Shastri2021,Wetzstein2020,Farmakidis2024,Markovic2020,Ahmed2025,Yin2025} which may not have been ideally suited to the von Neumann architecture but may find neuromorphic designs more energy efficient, faster and accessible \cite{Matuszewski2024,McMahon2023,Stroev2023}.

In parallel, several expected upcoming technologies rely on quantum features (quantum sensing \cite{Bongs_Bennett_Lohmann_2023}, communication \cite{Liu_2020}, and computing \cite{Takook2024}). As these often have optical components, it is important to characterize optical quantum states \cite{Schreiber25}, either for measurements during operation or for calibration purposes. This task is generally known as quantum state recognition \cite{Lvovsky2009}\ao{.} 

Quantum state recognition is a form of pattern recognition, often requiring the extraction of key features from a large data set (now states in the Hilbert space), which may very will be polluted with (hopefully irrelevant but still potentially distracting) errors due to the fragility of quantum states. Consequently, it is natural to consider the application of quantum versions of neural networks for such tasks\cite{ghosh2019quantum,Liu2025,Pehle2022,Biamonte2017}. 
In addition, although controlling quantum systems typically requires substantial effort and precision, the reservoir computing paradigm may offer a more flexible and efficient alternative. 

Reservoir computing \cite{GRIGORYEVAOrtega2018495,Jaeger2007} indeed offers a compelling advantage: it does not require tunability of the internal network. Instead, computation is performed by exploiting the intrinsic dynamics of a nonlinear system with complex, often random, internal couplings. Crucially, only the output layer is trained—typically via a simple linear regression on measured observables—thus avoiding the need for backpropagation or fine-tuning of the internal weights. This makes the approach highly appealing for physical platforms, including quantum or optical systems. Reservoir computing builds on the conceptual foundation of Echo State Networks and Liquid State Machines,
\cite{Jaeger2004,Schiller2005,Maass2002}
where a fixed recurrent neural network with a fading memory responds to inputs, and only the linear readout is optimized. Its successes underscore how rich, high-dimensional transient dynamics can be harnessed efficiently for machine learning tasks without requiring full control over the internal network structure \cite{Lukosevicius2012}.

Quantum Reservoir Computing -- using a quantum system as a reservoir in the above sense -- has, at least in principle, shown to be a promising approach to achieve a variety of quantum or classical tasks \cite{Nakajima_2020,Kobayashi24,Wang25,Mujal2023}, including those related to quantum state recognition 
\cite{ghosh2019quantum,Angelatos21}. 


In the simplest case, one aims to characterize a single quantum optical mode, with a specific but fixed spatial distribution. Also no further internal degrees of freedom are assumed. Typically, reservoir computing -- classical or quantum -- involves a network of nodes that are separated in real or sometimes reciprocal space \cite{VanderSandeBrunnerSoriano2017,Xu2020}.

Considering a single mode, one might naively expect that such patterns cannot exist. While a bosonic state can still have a non-trivial structure in the single-mode Hilbert space (e.g. represented in the Wigner phase space) \cite{Xu2021,Dudas2023} in principle, this is generally much harder to extract than average intensities.

In the present work, we find that a single mode (such as, e.g in a cavity) can still operate in a reservoir while remaining true to the original spirit of simply considering occupations (intensities) of different modes. We do this by exploiting the emission spectrum, which is nothing but the occupation at different frequency components. In particular, the nonlinear response of the optical cavity remaps the squeezing of an input state to signatures in the frequency domain.
In this way, the frequency distribution corresponds itself to a set of coupled ``reservoir'' nodes, which can be defined by representing the frequency response in given basis. The frequency spectrum gives a set of classical variables, which we can learn to extract the initial squeezing from.

Here, we focus on a setup consisting of a single nonlinear spatial mode or ``cavity'' for short. We will focus on the common case of $\chi^{(3)}$-nonlinearity, where the total interaction energy for $n$ photons in the mode scales as $\frac{U}{2}n^2$, leading to a blueshift $Un$ per photon.

Such a nonlinear spatial mode can appear in a variety of situations, for different ratios of the nonlinearity with respect to the interaction with the environment. Traditional (Kerr) nonlinear optics \cite{Boyd_2020} is a well-established field, with many successes including applications to neuromorphic computing, typically in the context of classical phenomena. At the other extreme end, superconducting circuits \cite{Blais2021}, as well as waveguide structures coupled to atoms \cite{Sheremet23}, have emerged as alternative platforms formally equivalent to cavities, with strong nonlinearities that enable access to the quantum regime but require stringent experimental conditions. In the intermediate regime, polaritonics considers optical photons hybridized with matter excitations, such as excitons in a semiconductor quantum well \cite{Kavokin2022}. These platforms are simultaneously being pushed toward more accessible conditions, such as room temperature \cite{Ghosh22,Wei2022}, while also exploiting higher nonlinearities to achieve quantum phenomena \cite{Liew23}. In the following, ``photon'' may also refer to any hybridized polaritons with a strong photon component \cite{Kalt_Klingshirn_2019}.

In this context, the spectrum of a single mode may provide route to neuromorphic computing that is simple and robust, features that are particularly relevant in the noisy intermediate-scale quantum era \cite{Deutsch2020}.

Building on these considerations, we now formalize how a single nonlinear optical mode can serve as a reservoir in the framework of quantum reservoir computing. Rather than relying on a spatial network of nodes, we interpret the different frequency components of the emission spectrum as effective nodes whose couplings arise from the intrinsic nonlinearity of the cavity. In the following section, we introduce the theoretical model describing this single-mode nonlinear cavity and derive how its emission spectrum encodes the necessary internal dynamics for computation.






\section{Model}

\begin{figure}
    \centering
    \includegraphics[width=\linewidth]{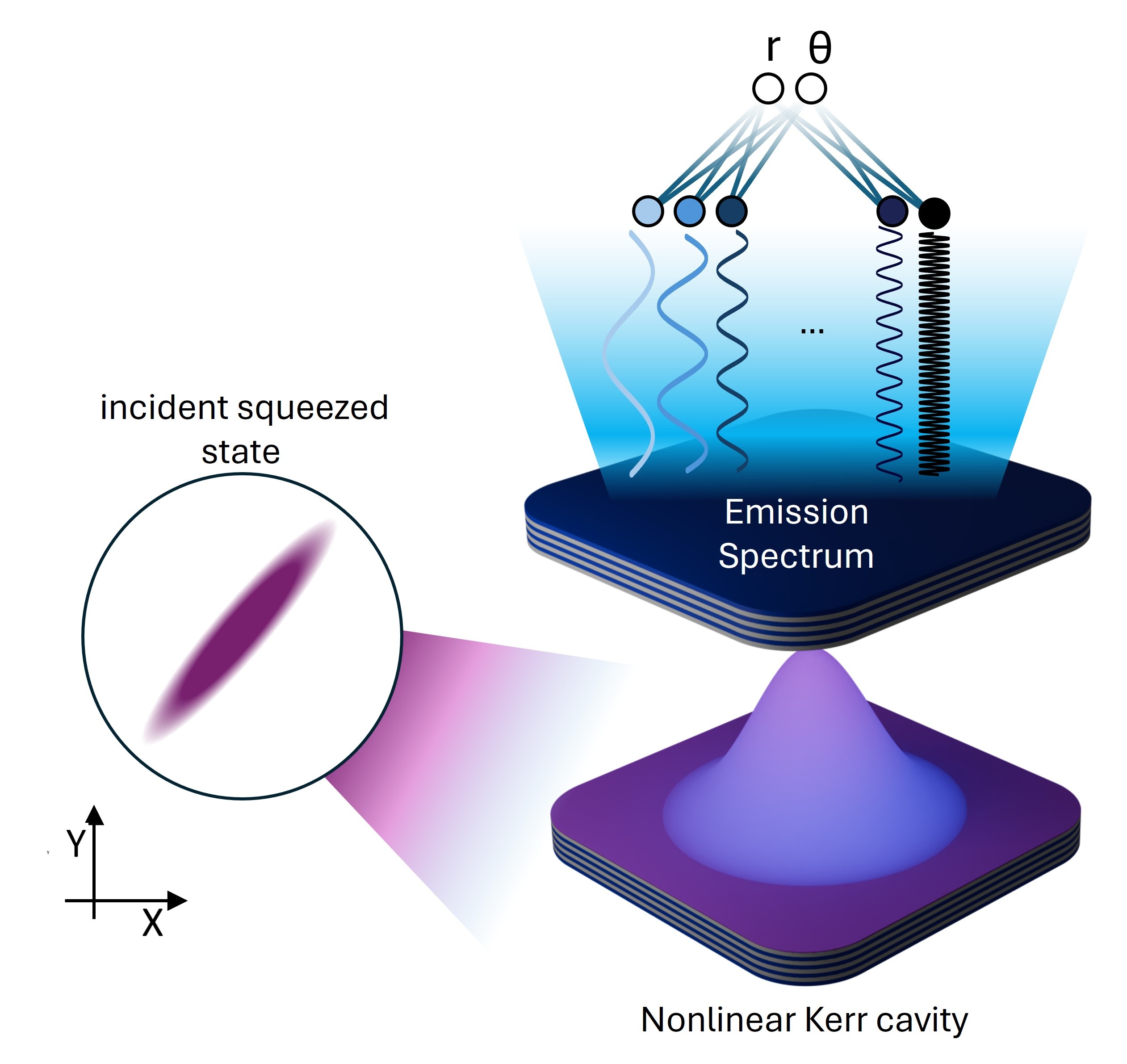}
    \caption{\textbf{Setup:} Quantum-optical squeezed states are continuously injected in a single-mode nonlinear cavity. The cavity's output radiation is then sent to a spectrometer, and the resulting parameters are then learned from the resulting emission spectrum}
    \label{fig:Setup}
\end{figure}

 In traditional reservoir computing, the input signal is spread over spatially separated but randomly interconnected discrete network nodes. In a photonic realization, such nodes might correspond to a lattice of multiple cavities or structurally localised photonic states in a single cavity, such as in a disorder potential \cite{Ballarini2020,Gan2025}. Interconnected waveguides realise conceptually similar setups \cite{VanderSandeBrunnerSoriano2017}.

 In contrast to previous approaches, the ``reservoir'' in this work relies on only one nonlinear optical mode, easily achievable in a single mode optical cavity (Fig. \ref{fig:Setup}). Generically, this is defined (in the frame rotating at the optical frequency--we will consider the case of the cavity and its external driving to be on resonance) by the Hamiltonian

\begin{equation}
    \hat{H}=\frac{U}{2}\hat{a}^\dagger\hat{a}^\dagger\hat{a}\hat{a},
\end{equation}

where $\hat{a}$ is the reservoir node bosonic annihilation operator\footnote{In this work, we use the term ``reservoir'' only in the sense of reservoir computing, unlike in the context of open quantum systems where it can denote a large environment or polaritonics where it denotes an ensemble of hot excitons.}.

The evolution of a quantum system that is, under appropriate assumptions \cite{Breuer_Petruccione_2010}, weakly coupled to an environment is described by the Lindblad master equation for the density matrix $\hat{\rho}$,

\begin{align}\label{eq:meq}
\partial_t\hat{\rho}&=\mathcal{L}\hat{\rho}=\mathcal{L}_{\hat{H}} \hat{\rho}+\sum_{\{\hat{L}\}} \mathcal{L}[\hat{L}]\hat{\rho}\\ &=-\frac{i}{\hbar}\comm{\hat{H}}{\hat{\rho}}+\sum_{\{\hat{L}\}}\left(-\frac{1}{2}\acomm{\hat{L}^\dagger\hat{L}}{\hat{\rho}}+\hat{L}\hat{\rho}\hat{L}^\dagger\right)
\end{align}

where curly braces denote the anticommutator and  $\mathcal{L}$ is the Liouvillian superoperator, the linear map describing the evolution of the density matrix. Lindblad operators ${\{\hat{L}\}}$ describe the interaction of the mode with the environment: photon losses at rate $\gamma_C$ are captured by $\hat{L}_l=\sqrt{\gamma_c}\hat{a}$. Continuous injection of squeezed states can further be described as coupling to a squeezed environment with an additional Lindblad operator as detailed in Sec \ref{subsec:squeezedenvmodel}, a framework that in fact generalizes the traditional description of coherent (resonant or quasi-resonant) driving.

Commonly, the time evolution is studied by solving master equation \ref{eq:meq} numerically as a set of ordinary differential equations. For the current single-mode case, relatively long time evolutions are needed while memory considerations don't pose such as stringent restriction, so that a more analytic approach is appropriate \footnote{In a state Hilbert space of dimension D, the size of a density matrix is $D^2$, which can already become challenging to save in computer memory for multimode bosonic systems. $\mathcal{L}$ has size $D^4$.}. By expressing $\mathcal{L}$ in matrix form, the propagator $e^{\mathcal{L}\Delta t}$ that describes the evolution over a time window $\Delta t$ is algebraically obtained \cite{Navarrete2015,Prosen_2008}. The propagator over any arbitrary integer multiple $t$ of $\Delta t$ is then given by $\left(e^{\mathcal{L}\Delta t}\right)^{t/\Delta t}$. After the steady-state is reached algebraically  as $\hat{\rho}_{SS}= e^{\mathcal{L}t_0}\ket{0}\bra{0}$, we compute the first-order coherence function

\begin{equation}
\ev{\hat{a}^\dagger(\tau)\hat{a}}=\Tr(\hat{a}^\dagger e^{\mathcal{L}\tau} \hat{a}\hat{\rho}_{SS})\propto g^{(1)}(\tau)    
\end{equation}

The emission spectrum is then obtained as the Fourier transform of this quantity \cite{Landi2024}:

\begin{align}
S(\omega)&\propto\ev{\hat{a}_{\omega}^\dagger\hat{a}_{\omega}}\\
&\propto\int_{-\infty}^{\infty} \ev{\hat{a}^\dagger(\tau)\hat{a}}e^{-i\omega\tau}\dd{\tau}\\&\propto\Re\int_0^{\infty} \ev{\hat{a}^\dagger(\tau)\hat{a}}e^{-i\omega\tau}\dd{\tau},\label{eq:specintegral}    
\end{align}

where constant proportionality factors are unimportant.
To define the nodes of the reservoir network, we start from $S(\omega)$, a continuous set of variables representing the occupation at frequency $\omega$. A singular contribution $\sim\delta(\omega)$, resulting from the coherent field of the average displacement, is removed. A definition of discrete network nodes can also be recovered by defining $m$-moments of the spectrum, via $\tilde{M}_m=\int S(\omega) \omega^m \dd{\omega}$ and, with normalization of the higher moments,$M_0=\tilde{M}_0,\quad M_m=\tilde{M}_m/M_0)\quad (m\neq 0)$ for $m\in \{0,1,\ldots,4\}$. The presence of the nonlinearity couples the frequency components $S(\omega)$, and as such also the moments $M_m$. 
Through this nontrivial interdependence, they effectively act as coupled reservoir nodes. In the reservoir computing framework, it is not necessary to identify this interdependence explicitly; it is sufficient that it exists in principle.

With the values of $M_m$, a linear regression is then performed with the parameters characterizing the squeezed input states for the ``training'' of the network (or further machine learning techniques, see subsec. \ref{subsec:multivar}). Interestingly, the spectra of a coherently driven nonlinear oscillator are trivial in the mean-field, but become rich because of fluctuations including those of quantum origin \cite{Drummond_1980}.

Within this general framework, we demonstrate that our system can recognize a squeezed light distribution 
that is continuously injected from the environment, agnostic to the origin of the squeezed state.

We will then apply this to a concrete setup in the form of a realistic OPO \cite{Burks09} driving a planar exciton-polariton microcavity \cite{Kavokin_Baumberg_Malpuech_Laussy_2017,Carusotto2013} in Subsec. \ref{subsec:OPOresults}, where the microcavity photons hybridize with quantum well excitons. This model corresponds to an example of  complete representation of an experimental implementation, where the source of the squeezing and the potential backaction of our system on it is explicitly accounted for. It should be noted that although we consider an exciton-polariton system with realistic material parameters for the sake of concreteness, the scheme is in principle applicable to any driven-dissipative nonlinear spatial mode.

\section{Results and discussion}

\subsection{Emission spectra of squeezed reservoirs \label{subsec:squeezspectra}}

\begin{figure*}
    \centering
    \includegraphics[width=\linewidth]{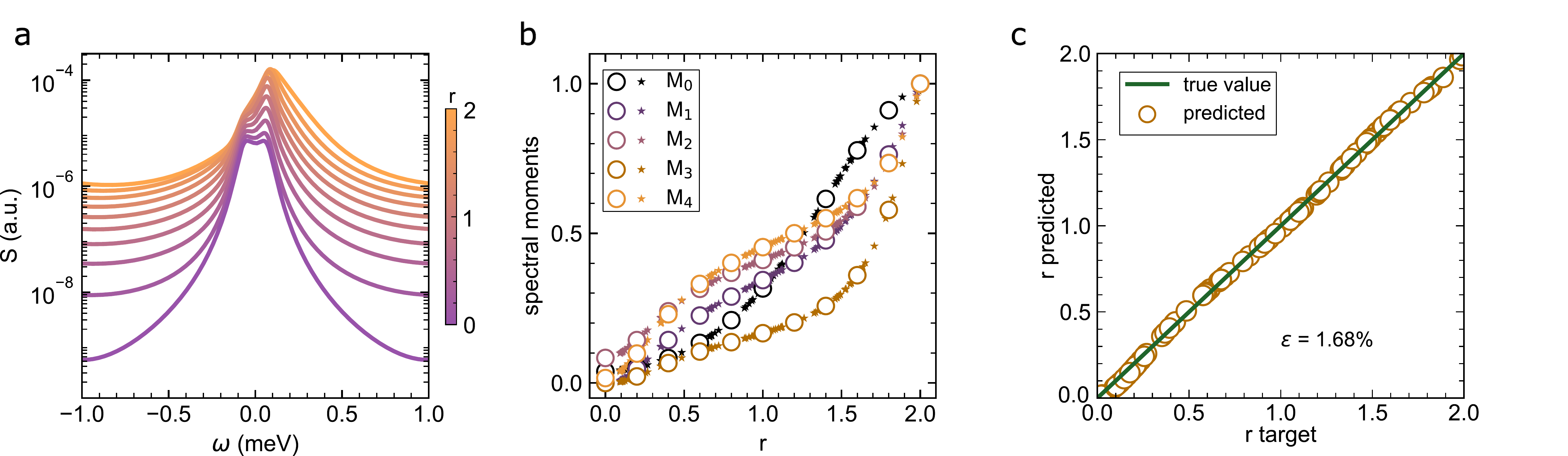}
    \caption{
    \textbf{Learning squeezing strength} (a): Spectra of a polariton microcavity coupled to a squeezed environment, for different values of squeezing strength $r$ at $\theta=0, \alpha_D=5,\Bar{n}=0$ (b):
    moments $M_m$ as a function of $r$ for the same training data as the left panel (circles) and 100 random testing data in the same range (stars). The moments are rescaled to their respective maximum values for display (c): individual comparison of actual values of $r$ versus the predicted on testing data . This leads to a NRMSE of $\varepsilon\approx1.7\%$.
    Parameters $U=12~\mu eV$ \cite{Kasprzak2007,Liew2010}  and $\gamma_c= 67~ns^{-1}=44.1~\mu eV$.}
     \label{fig:rdepspectra}
\end{figure*}

We consider squeezed states parametrized by a complex squeezing parameter $\zeta=re^{2i\theta}$ where the squeezing strength $r\in [0,2]$ and the squeezing angle $\theta$ are real-valued and in an experimentally accessible range. Furthermore, the states will be displaced from the origin by a mean-field value $\alpha_D$ (chosen to be real and positive without loss of generality), as well as thermally broadened with average occupation $\Bar{n}$. Such a type of states can be written in the following form
\begin{equation}
   \hat{\rho}=\hat{D}(\alpha_D) \hat{S}(\zeta) \hat{\rho}_{th} \hat{S}(\zeta)^\dagger \hat{D} (\alpha_D)^\dagger, 
\end{equation}
where the density matrix operator of the thermal state is given by $\hat{\rho}_{th}=\sum_n \frac{\Bar{n}^n}{(\Bar{n}+1)^{n+1}}\ket{n}\bra{n}$, the displacement operator by $\hat{D}(\alpha_D)=e^{\alpha_D\hat{a}^\dagger-\alpha_D^*\hat{a}}$ and squeezing operator is defined as $\hat{S}=e^{\frac{-\zeta}{2}\hat{a}^{\dagger 2}+\frac{\zeta^*}{2}\hat{a}^2}$ \cite{Walls_Milburn_2008}.

In Figure \ref{fig:rdepspectra}a, we show how the spectra $S(\omega)$ vary with $r$ at constant $\theta, \alpha_D, \Bar{n}$, as do their corresponding moments $M_m$ (panel b, circles). Having established the presence of a significant response to $r$, we proceed to utilize it for reservoir computing, where these $N_\mathrm{train}=10$ equidistant points are used as training data.

A linear regression\footnote{Generally, a ridge regression is used for reservoir computing, but we found that the ridge parameter can be chosen vanishingly small in this particular case.} is performed on these to train the model. $N_\mathrm{test}=100$ other squeezed states, with random, uniformly distributed, $r\in [0,2]$ are also displayed in panel b along with their moments $M_m$ (stars), that will be used to test the model. In panel c, we plot the true (target) versus the predicted values of $r$ for all these data points.The strong agreement between them demonstrates the accuracy of the prediction.  As a figure of merit for the whole dataset, we utilise the Normalised root-mean-square error (NRMSE) $\varepsilon=\sqrt{\ev{(y-y_t)^2}/\ev{y_t^2}}$. where $y$ and $y_t$ denote the predicted and target parameter values, respectively. In the aforementioned case with $y=r$, a value $\varepsilon\approx 1.7\%$ is achieved.

In Subsec \ref{subsec:noiseandmoments}, the use of moments $M_0 \ldots M_4$ is validated further. On the one hand, this approach approaches the precision of using the whole spectrum efficiently (Fig. \ref{fig:Momentscandprediction}), while it is also rather robust to the presence of noise (Fig. \ref{fig:noisedependence}.)

The same analysis is performed for all aforementioned parameters of the squeezing, with results summarized in table \ref{tab:performance}. As we see, $\varepsilon$ is of the order of only a few percent, highlighting the suitability of this approach.

\begin{table}
    \centering
    \begin{tabular}{ccc}
        Variable $y$  & Conditions & $\varepsilon$ \\
        \hline
        $r\in [0,2]$ & $\alpha_D=5, \theta=0,\Bar{n}=0$  & $1.7\%$\\
         $r\in [0,2]$& $\alpha_D=5, \theta=\pi/2,\Bar{n}=0$ & $2.5\%$\\
         $\theta\in [0,\pi/2]$& $\alpha_D=5, r=2,\Bar{n}=0$  & $0.32\%$ \\
        $\alpha_D\in [0,5]$ &$r=2, \theta=0,\Bar{n}=0$  & $1.8\%$ \\
        $\alpha_D\in [0,5]$ & $r=2, \theta=\pi/2,\Bar{n}=0$ & $2.4\%$\\
        $\Bar{n} \in [0,10]$ & $\alpha_D=5, \theta=0,r=0$ & $10.1\%$\\
    \end{tabular}
    \caption{\textbf{Performance on different parameters} overview of the error $\varepsilon$ after training/testing different parameters characterizing Gaussian states: squeezing $r$, angle $\theta$, displacement $\alpha_D$, under otherwise same conditions as in Fig. 1.}
    \label{tab:performance}
\end{table}

These results confirm that the emission spectrum of a single nonlinear cavity can reliably encode the parameters of the squeezed input state. Having established this basic recognition capability, we next examine the role of the system’s intrinsic nonlinearity U, which governs the degree of coupling between the spectral components—that is, between the effective nodes of the reservoir. Understanding how the performance scales with U is essential for assessing both the expressiveness and the physical limits of this approach.

\subsection{Dependence on the Nonlinearity}

The expressiveness that allows neural networks to express arbitrary functions  requires nonlinear elements \cite{Fu2024}. In the reservoir computing framework, this nonlinearity is fixed and ingrained in the reservoir. 

\begin{figure}
    \centering
    \includegraphics[width=\linewidth]{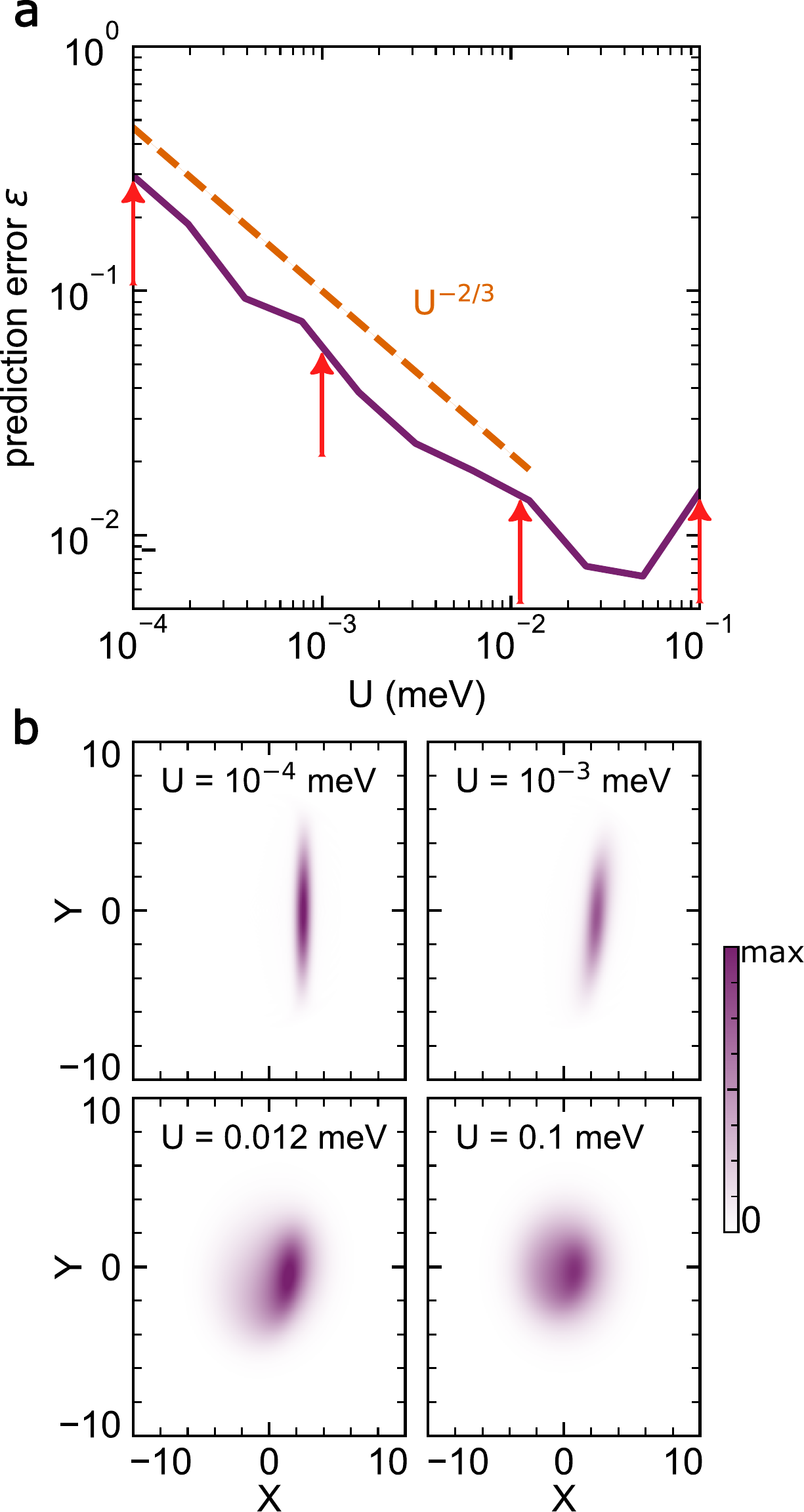}
    \caption{\textbf{Dependence on reservoir nonlinearity} (a): the prediction error on $r$ (same task as in Fig. \ref{fig:rdepspectra}) decreases with $U$, as is well captured by a power law decay $\varepsilon\propto U^{-2/3}$. At higher values of $U$, the performance saturates. (b): Wigner-functions in optical phase space of the steady-states in the cavity for the different values of $U$ that are indicated by the red arrows in (a). At the lowest values $U= 10^{-4}~meV$ the nonlinearity has little effect and the microcavity steady state is directly proportional to the squeezed input state. For increasing $U$, the behavior of the state becomes progressively richer while $\varepsilon$ decreases. At $U=0.1~meV$ the squeezing becomes suppressed, to which we can attribute the saturation in performance. The quadrature variables in the reservoir mode are defined as  $\hat{X}=\frac{\hat{a}+\hat{a}^\dagger}{2}$ and $\hat{Y}=\frac{\hat{a}-\hat{a}^\dagger}{2i}$. }
    \label{fig:Udependence}
\end{figure}

While a nonlinear readout applied to an otherwise linear system could, in principle, play a role in this behavior\cite{Govia21,Neill2025} we find that the dominant nonlinear mechanism arises from the physical photon-photon interaction strength $U$.
To study this, the same task of Fig. \ref{fig:rdepspectra} was repeated across a range of different values of $U$. In figure \ref{fig:Udependence}a, we indeed find that the performance depends strongly on this nonlinearity. We observe a clear power-law reduction on the NRMSE with nonlinearity that is fitted as $\varepsilon\propto U^{-2/3}$, towards the experimental value $U=12~\mu eV$ (which was used elsewhere in this work). 

At even higher values of $U$, the error saturates however. To understand the source of this behaviour, we study the Wigner functions of the steady state solutions for different values of $U$ in the right panels of Fig.~\ref{fig:Udependence}b.

For moderate values of $U$, the system exhibits a non-trivial response to the squeezed states, whereas higher values of $U$ substantially  suppress the squeezing, thereby reducing sensitivity.


\subsection{Generalisation of learning\label{subsec:generalisation}}

\begin{figure}
    \centering
    \includegraphics[width=\linewidth]{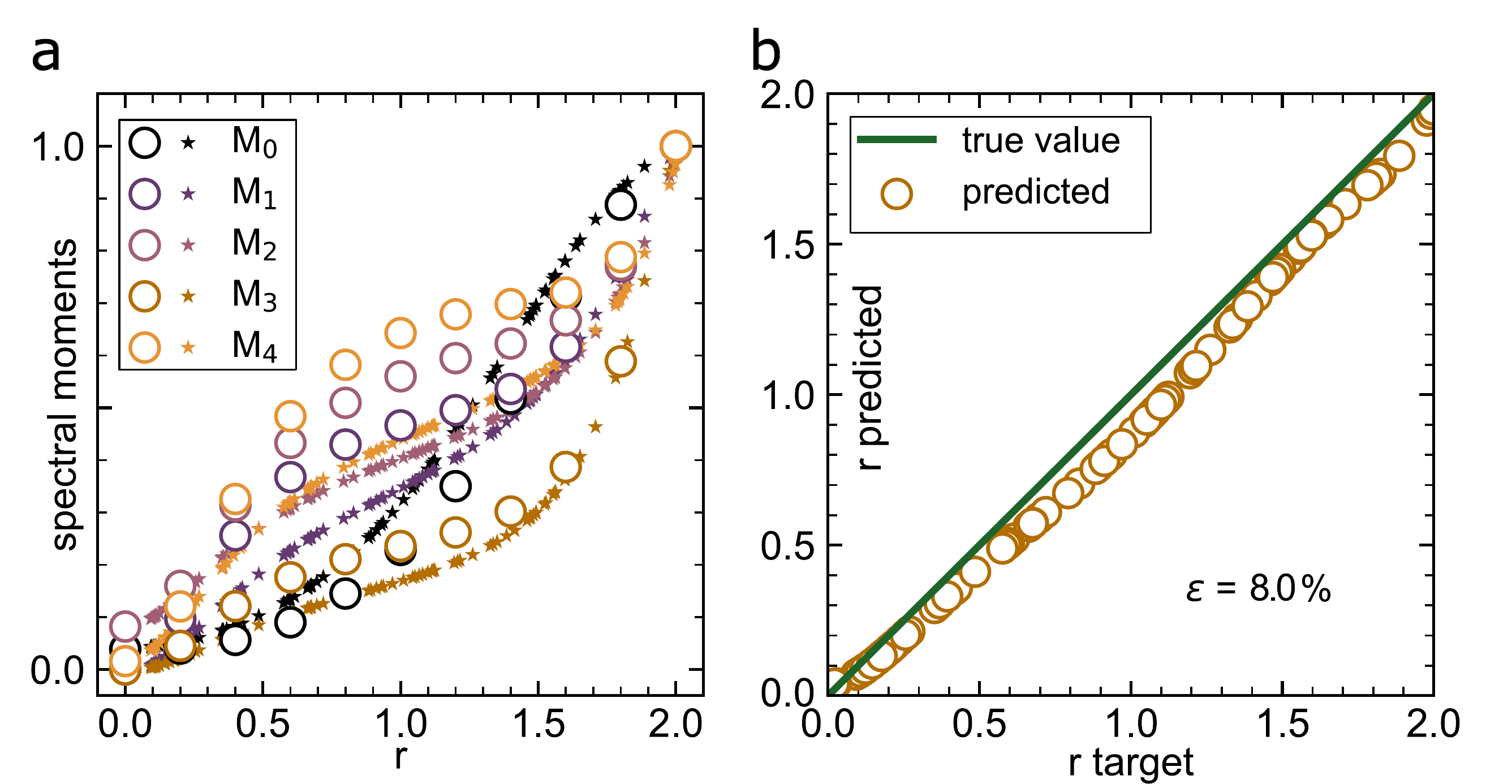}
    \caption{\textbf{Generalisability of learning:} rescaled moments $M_m$ as function of $r$, when training is performed at $\theta=\pi/2$ (circles) and testing at $\theta=0$ (stars). The prediction error on $r$ is still relatively small at $\varepsilon=8.0\%$, highlighting that learning in the network generalises.}
    \label{fig:generalisation}
\end{figure}

As a hallmark of quantum reservoir performance, one may assess its generalization capability—the ability to operate accurately on data distinct from the training set. Fig \ref{fig:generalisation} presents an example of such behavior. In this scenario, the network is trained to identify $r$ as in \ref{fig:rdepspectra}, with training performed at $\theta=\pi/2$ (phase squeezing) and testing at $\theta=0$ (number squeezing). A satisfactory performance of $\varepsilon\approx8\%$ is still obtained.

\subsection{Multivariate predictions with advanced Neural networks \label{subsec:multivar}}
\begin{figure*}
    \centering
    \includegraphics[width=\linewidth]{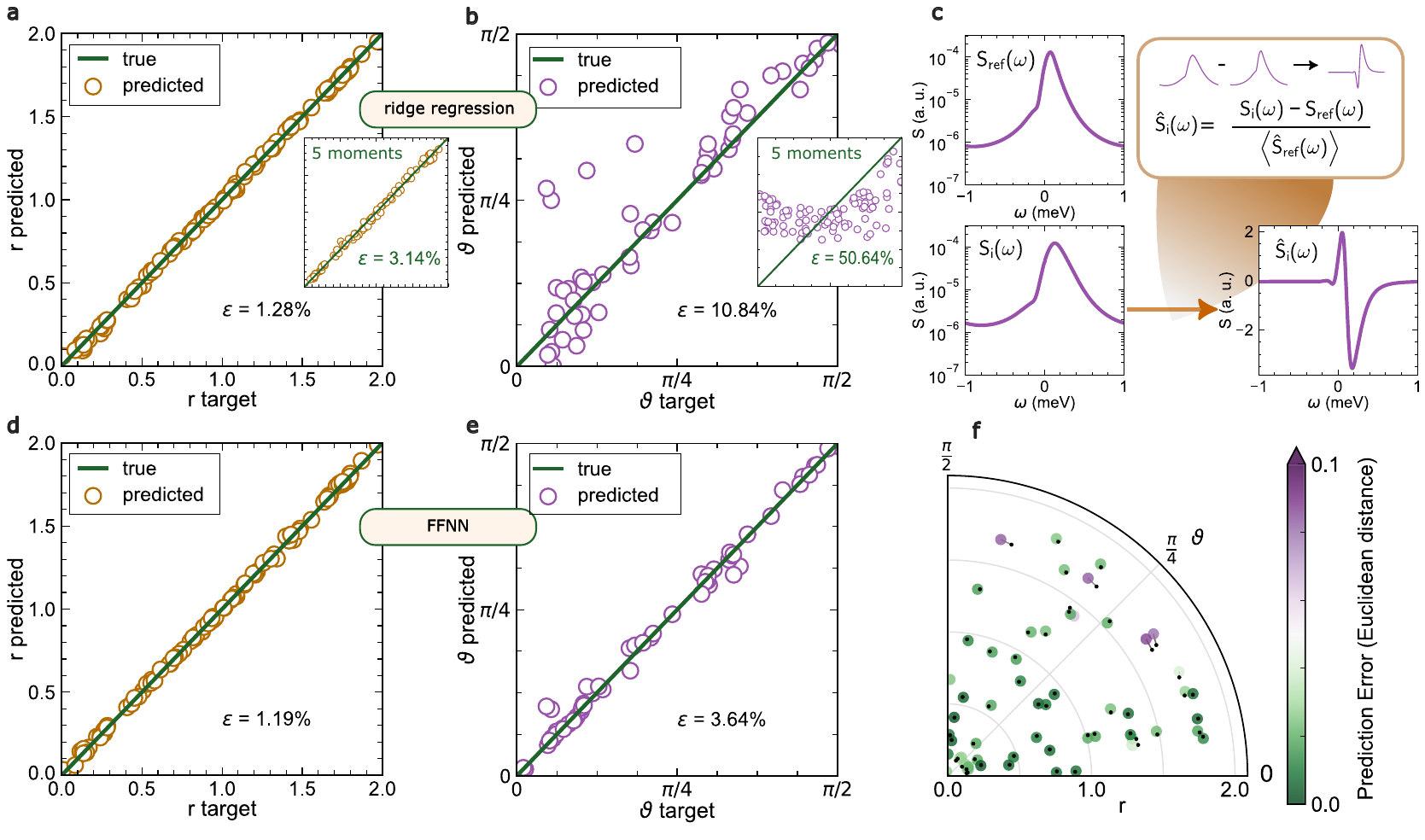}
    \caption{\textbf{Multivariate prediction with a hybrid quantum-classical model.} In a dataset with variability both in $r$ and $\theta$, learning with moments $M_0-M_4$ as before works well to predict $r$ (inset of panel \textbf{(a)}), but less for $\theta$ (inset of \textbf{(b)}). Results improve significantly by training on the pre-processed (see panel \textbf{c}) full spectrum, as depicted on the main panels of \textbf{(a,b)}. Further improvement is obtained using a feedforward neural network model trained on the pre-processed spectral data (panels \textbf{d,e}).
    Panel \textbf{c} depicts the renormalization procedure. The differences $\hat{S}_{i}(\omega)$ obtained by subtracting a reference spectrum $S_{ref}(\omega)$ from each of the spectra $S_{i}(\omega)$ in the dataset are renormalized (divided by the mean over $\omega$ $\left\langle S_{i}(\omega)\right\rangle$) and used as training examples for training the neural network. Panel \textbf{f} depicts a polar coordinate representation of the predicted data, with colored points corresponding to predicted squeezing parameters, and black points to the target values. The color of the predicted points corresponds to the prediction error (measured as the Euclidean distance $d=\sqrt{r_p^2+r_t^2-2r_pr_t\cos{(\theta_p-\theta_t})}$ between predicted $(r_p, \theta_p)$ and target $(r_t,\theta_t)$ squeezing parameters).}
     \label{fig:enhanced}
\end{figure*}

We now move on to the multivariate case. In particular, we will focus on the case where $r,\theta$ are varying under fixed $\alpha_D, \Bar{n}$. 800 squeezed states with random $r \in [0, 2]$ and $\theta \in [0,\pi/2]$ are generated, and divided in $N_{\text{train}} = 700$ training states and $N_{\text{test}} = 100$ states to test. As we see in \ref{fig:enhanced} a,b; insets, the approach from the preceding sections using $M_0-M_4$ leads to an adquate prediction of $r$ ($\varepsilon_r\approx 1.3 \%$), but not $\theta$ ($\varepsilon_\theta\approx 51\%$).  We then reconsider the full (pre-processed, see panel c and subsec \ref{subsec:multimet}) spectra.
Using the full spectrum $S(\omega)$ leads to a significant improvement (a,b; main panels), with $\varepsilon_\theta\approx11\%$.

Still, the performance is lower than that of the single-variable scenario in Table \ref{tab:performance}, motivating a deeper investigation.

We employed a feedforward neural network (FFNN) consisting of three fully connected (dense) layers with rectified linear unit (ReLU) neuron activation functions (implemented on the digital computer used for post-processing). The number of neurons per layer was set as $N_{1} = 200$,  $N_{2} = 200$,  $N_{3} = 100$, and the output layer consisted of two neurons with linear activation functions. The model was trained using the \textit{Adam} optimizer with a learning rate of $\eta = 5 \times 10^{-5}$, the loss function was \textit{Mean Squared Error} 
(equivalent to the NRMSE as a monotonic function), and the model took 2500 epochs to converge with batch size of 100 states. The predictions of the FFNN are very precise for both parameters ($\varepsilon_{r} = 1.19 \%, \varepsilon_{\theta} = 3.64 \%$).
While the additional neural network has been shown to dramatically decrease the prediction errors in the multivariate case, 700 training examples have been used. This seems significantly more than the number of training examples needed to predict only one of the parameters with the use of linear regression in the preceding sections. However, we note that 10 equidistant training points for a single variable, would correspond to a regular grid of 100 training points in a two-variable case. In the multivariate case considered here, the 700 training points were randomly selected instead. We have verified that also in this case, the physical nonlinearity $U$ is essential for the neural network to work.

\subsection{A realistic slow Optical Parametric Oscillator \label{subsec:OPOresults}}

The coupling mechanism discussed so far is quite general and applies to  the continuous driving by single-mode squeezed states. However, realistic sources are rarely truly single-mode \cite{Bachor_Ralph_2019,Walls_Milburn_2008} and, in particular, may have their own complex spectral dependence. To study if this difference is critical for the working of the setup, we proceed to study the reservoir computing approach of an actual, degenerate Optical Parametric Oscillator (OPO).
We use parameters of a well characterized OPO that is naturally compatible with atomic memories \cite{Burks09}. For the reservoir mode, a GaAs exciton-polariton microcavity is used, with the same parameters as the reservoir mode of the preceding sections. As these platforms for the source and the reservoir cavity are very different and have seemingly incompatible timescales, it is a highly non-trivial question if such reservoir approach will work.

Following Ref. \cite{Walls_Milburn_2008}, the Hamiltonian of an OPO below threshold is given by

\begin{equation}
    \hat{H}_\text{OPO}=\frac{i}{2}\left(G\hat{b}^{\dagger2}-G^*\hat{b}^2\right), \label{eq:OPOham}
\end{equation}
where $G$ is the pump beam approximated by a classical value. It is further subject to additional losses through the Lindblad operator $\sqrt{\gamma_s}\hat{b}$. The steady state of such OPO is a squeezed, mixed vacuum that interpolates between the plain vacuum at $G=0$ and diverging squeezing at the threshold $G_{th}=\frac{\gamma_s}{2}$. The squeezed vacuum of the OPO is then mixed with a coherent state on a beamsplitter to achieve a displaced squeezed state. The latter couples to the microcavity through the cascaded coupling scheme \cite{gardiner2004quantum,Carreno2016,Gardiner1993,Carmichael1993}. Due to the higher complexity required of solving the combined source+reservoir system, the Positive P phase space method \cite{Deuar2021} forms an appropriate tool, whereas the separation of timescales ($\gamma_s=10~ms^{-1} \ll \gamma_c=67~ns^{-1}$) warrants a quasi-static approximation. Refer to subsec \ref{subsec:OPOmodel} for further details on the computation of these spectra.





\begin{figure}
    \centering
    \includegraphics[width=\linewidth]{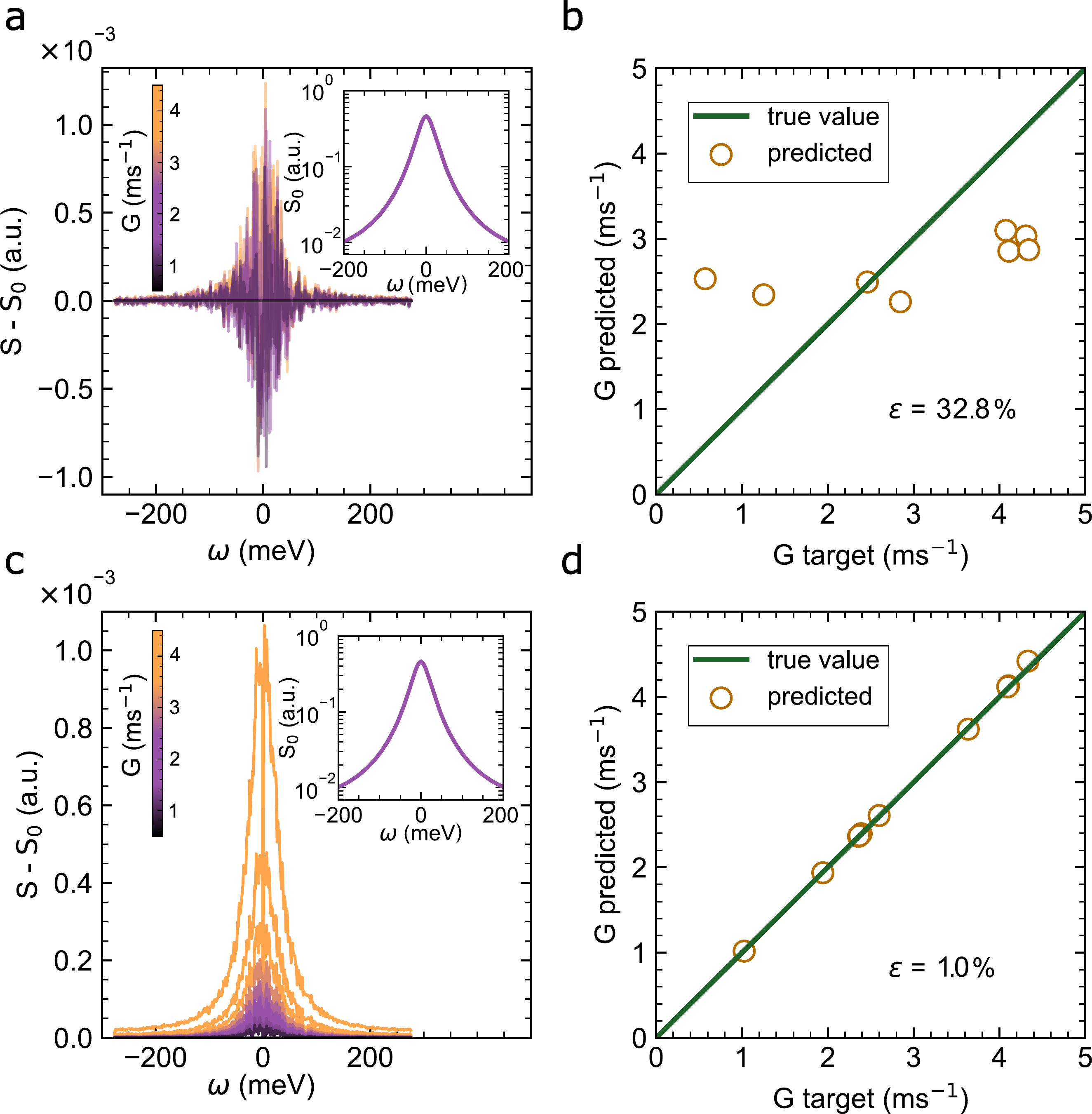}
    \caption{\textbf{Spectroscopy reservoir approach, applied to a realistic OPO:} a) Spectral differences of the polaritons driven by an OPO for different values of $G$ with respect to the unsqueezed ($G=0$)  coherently pumped spectrum $S_0$, whole spectra in the inset. Spectra for test data $G=0~ms^{-1},0.5~ms^{-1},\ldots,4.5~ms^{-1}$ are shown from purple to yellow. Moments are extracted from the spectral difference. Recognizing $G$ on 10 random test data values with $G$ in the same range results in prediction error  $\varepsilon=32.8\%<\varepsilon_b$, below the baseline (always using the average value $G_{avg}=2.25$ as an estimate would yield $\varepsilon_b=50\%$). Panel b) displays the prediction for the individual data points, and suggests that the highest values of $G$  are recognized more accurately. The aforementioned result is still limited by statistical noise from finite evolution in the simulation (see main text and \ref{subsec:OPOmodel} for details). In c,d) we see the potential performance when this statistical noise is reduced (here simulated by returning the same random number stream to the simulations at different $G$ values), reaching $\varepsilon=1.0\%$.}
    \label{fig:OPO}
\end{figure}

With these spectra, we proceed to the task of learning the value of $G$ by means of the moments $M_0-M_4$ by the approach of sections \ref{subsec:squeezspectra}-\ref{subsec:generalisation}. $N_\mathrm{train}=10$ training spectra at $\frac{G}{G_\mathrm{th}}=0,0.1,\ldots,0.9$ are plotted on \ref{fig:OPO}a (inset). As the spectrum is strongly dominated by the one of the fixed coherent displacement (which exactly equals $S_0(\omega)$, the spectrum at zero squeezing $G=0$), we focus on the moments of $S(\omega)-S_0(\omega)$, plotted on the main panel. A linear regression is performed between $G$ and the moments of these spectral differences to train the reservoir. The model is than tested for $N_\mathrm{test}=10$ test spectra with random $0\leq G \leq 0.9 G_\mathrm{th}$. In panel \ref{fig:OPO}b, we see that the prediction of these values of $G$ leads to large deviations, as through a large range of $G$, the target and predicted values are not very correlated. At the highest values of $G$,  such a correlation is picked up though. Overall, this leads to a prediction NRMSE of $\varepsilon\approx33\%$. Although this deviation is substantial, it still significantly surpasses (is smaller than) the baseline value $\varepsilon_b=50\%$ corresponding to always predicting the average value $G=0.45 G_\mathrm{th}$.

We note however, that due to numerical constraints, the simulated spectra were only obtained from a numerical evolution time that would correspond to a physical time somewhere between $3~\mu s$ and $3~ms$ (see \ref{subsec:OPOmodel} for details), that results in a low signal-to-noise ratio for the spectra, as can be readily seen on panel a. Only the squeezing spectra for the highest values of $G$ rise above the noise in this case. 

To obtain an idea of the capabilities of the setup with longer exposure times (still well within practical constraints for experiments), we eliminate the effect of the variability of the noise
between the different spectra. This is achieved by repeating the analysis, while using the same noise realisation (random number stream in the numerical simulations) for each spectrum.

In \ref{fig:OPO}c, we observe indeed that the spectra for different $G$ are easier to discern like that. This is confirmed in \ref{fig:OPO}d, where we now observe a strong correlation between target and predicted values of $G$. The NRMSE here reaches $\varepsilon\approx 1\%$, highlighting the potential of the approach.

Altogether, these results demonstrate that a single nonlinear mode—when driven by either idealized squeezed states or realistic sources such as an OPO—can perform efficient quantum state recognition based solely on its spectral response. To ensure the reproducibility and transparency of our analysis, we now detail the underlying theoretical framework, numerical implementations, and data-processing procedures used throughout this work.


\section{Methods}

\subsection{Squeezed environment \label{subsec:squeezedenvmodel}}

A typical approach to driving by quantum-states is by the so called cascaded coupling mechanism, \cite{gardiner2004quantum,Carreno2016,Gardiner1993,Carmichael1993} and has been used in the description of quantum reservoir computing before \cite{ghosh2019quantum}. However, in this formalism, losses of the source term have to be also considered to form a physical model, making it unsuitable for static sources (if not, the combined source–quantum reservoir system does not admit a Lindblad representation and thus does not generate a completely positive, trace-preserving map for the density matrix). Continuous driving can only be described by the cascaded coupling formalism if the whole dynamics of state generation in the source is incorporated explicitly to offset the losses. Such a specific case is considered in sec. \ref{subsec:OPOmodel}.

In the current section, we devise an alternative approach to describe continuous pumping with squeezed states as coupling to a squeezed and displaced environment, without further assumptions on the origin of this squeezing.

When a cavity is coupled to the (plain) vacuum as its environment, it will cause a decay (photon losses) according to the Lindblad operator $\sqrt{\gamma}\hat{a}$. If the environment has a thermal occupation, also incoherent gain $\sqrt{P}\hat{a}^\dagger$ is present as a second lindblad operator, with the ratio $P/\gamma$ set by the temperature. 

These Lindblad operators have microscopic origin in a hopping where the modes $\hat{a}$ in the cavity are coupled to the modes $\hat{a}_E$ in the environment \cite{Breuer_Petruccione_2010}.

However, also a coherent pump can be considered in this framework, as decay into a displaced vacuum by Lindblad operator $\Lop_d=\hat{D}(\alpha)\sqrt{\gamma}\hat{a}\hat{D}(\alpha)^\dagger=\sqrt{\gamma}(\hat{a}-\alpha)$. 

Indeed, we can prove that

\begin{align}
    \mathcal{L}[\Lop_d]\hat{\rho}&=\Lop_d\hat{\rho}\Lop_d^\dagger-\frac{1}{2}\acomm{\Lop_d^\dagger\Lop_d}{\hat{\rho}}\\
    &=-\gamma\comm{\frac{\alpha^*\hat{a}-\alpha\hat{a}^\dagger}{2}}{\hat\rho}+\mathcal{L}[\sqrt{\gamma}\hat{a}]\\
    &=\frac{-i}{\hbar}\comm{\hat{H}_p}{\hat{\rho}}+\mathcal{L}[\sqrt{\gamma}\hat{a}]\ao{,}
\end{align}
with $\hat{H}_p=\hbar\gamma \frac{\alpha^*\hat{a}-\alpha\hat{a}^\dagger}{2i}$, which is the Hamiltonian representing coherent driving with amplitude $F=\hbar\frac{\gamma}{2}\alpha$.

Here, we generalize this construction by considering coupling to an environment that is squeezed in addition to displaced \cite{Breuer_Petruccione_2010}. Like displacement (translation), squeezing and rotation form a symplectic/canonical symmetry transformation in phase space
\cite{Ferraro_Olivares_Paris_2005,Glauber91}.
To make this more explicit, consider an operator $\hat{O}$ with elements $\bra{m}\hat{O}\ket{n}$ in the original Fock basis. We want to find the operator $\hat{\Tilde{O}}$ that has the same representation in a squeezed, and displaced reference frame with basis $\ket{n,\zeta,\alpha}=\hat{D}(\alpha)\hat{S}(\zeta)\ket{n}$. Then

\begin{align}    \bra{m}\hat{O}\ket{n}&=\bra{m,\zeta,\alpha}\hat{\Tilde{O}}\ket{n,\zeta,\alpha}\\
&=\bra{m}\hat{S}^\dagger(\zeta)\hat{D}^\dagger(\alpha)\hat{\Tilde{O}}\hat{D}(\alpha)\hat{S}(\zeta)\ket{n},
\end{align}

which is achieved for $\hat{\Tilde{O}}=\hat{D}(\alpha)\hat{S}(\zeta)\hat{O}\hat{S}^\dagger(\zeta)\hat{D}^\dagger(\alpha)$.

Filling in $\hat{a},\hat{a}^\dagger$ for $\hat{O}$, we find the transformed annihilation and creation operator,

\begin{align}
    \hat{b}&=(\hat{a}-\alpha)\cosh{r}+e^{2i\theta}(\hat{a}^\dagger-\alpha^*)\sinh{r},\\
    \hat{b}^\dagger&=e^{-2i\theta}(\hat{a}-\alpha)\sinh{r}+(\hat{a}^\dagger-\alpha^*)\cosh{r},
\end{align}
which take the form of a Bogoliubov transformation.

Creating a pure Gaussian state in the cavity with squeezing $\zeta=re^{2i\theta}$ and displacement $\alpha$ can thus be done with the Lindblad operator $\sqrt{\kappa}\hat{b}$, where $\kappa$ is an arbitrary proportionality constant indicating the strength of the coupling (we will assume $\kappa=\gamma_c$). 

Thermal mixed states of average occupation $\Bar{n}$, in general, arise naturally as the steady state by an interplay of gain $\sqrt{P}\hat{a}^\dagger$ and losses $\sqrt{\gamma}\hat{a}$ in the system such that $P(\Bar{n}+1)=\gamma\Bar{n}$ where the ``+1'' on the LHS represents spontaneous emissions. Likewise, we can realize squeezed and displaced thermal states by introducing the additional Lindblad operator $\sqrt{R}\hat{b}^\dagger$ and setting
\begin{equation}
    R=\frac{\Bar{n}}{\Bar{n}+1}\kappa.
\end{equation}

In fact, displacement and squeezing decouple, the effect 

\begin{align}
    &\mathcal{L}[\sqrt{\kappa}\hat{b}]\hat{\rho}+\mathcal{L}[\sqrt{R}\hat{b}^\dagger]\hat{\rho}\nonumber\\
    &=\kappa\hat{b}\hat{\rho}\hat{b}^\dagger-\frac{\kappa}{2}\acomm{\hat{b}^\dagger\hat{b}}{\hat{\rho}}+R\hat{b}^\dagger\hat{\rho}\hat{b}-\frac{R}{2}\acomm{\hat{b}\hat{b}^\dagger}{\hat{\rho}}\\    
&=-i\comm{(\kappa-R)\frac{\alpha^*\hat{a}-\alpha\hat{a}^\dagger}{2i}}{\hat{\rho}}+\kappa\hat{\sigma}\hat{\rho}\hat{\sigma}^\dagger-\frac{\kappa}{2}\acomm{\hat{\sigma}^\dagger\hat{\sigma}}{\hat{\rho}}\nonumber\\&+R\hat{\sigma}^\dagger\hat{\rho}\hat{\sigma}-\frac{R}{2}\acomm{\hat{\sigma}\hat{\sigma}^\dagger}{\hat{\rho}}\\
&=\frac{-i}{\hbar}\comm{\hat{H}_p}{\hat{\rho}}+\mathcal{L}[\sqrt{\kappa}\hat{\sigma}]\hat{\rho}+\mathcal{L}[\sqrt{R}\hat{\sigma}^\dagger]\hat{\rho}\,
\end{align}

where 
\begin{equation}
 \hat{\sigma}=\hat{S}(\zeta)\hat{a}\hat{S}^\dagger(\zeta)=\hat{a}\cosh{r}+e^{2i\theta}\hat{a}^\dagger \sinh{r},   
\end{equation}
 and 
\begin{equation}
\hat{H}_p=iF\hat{a}^\dagger-iF^*\hat{a},    
\end{equation}
for $F=\hbar\frac{\kappa-R}{2}\alpha$.
The squeezing must then be modeled with the undisplaced Bogoliubov operators $\hat{\sigma}$ while $\hat{H}_p$ is part of the Hamiltonian.

Therefore, driving the system by a squeezed and displayed environment is equivalent to combining a coherent pump (represented by $\hat{H}_p$) with pure squeezed losses $\hat{\sigma}$. For thermally broadened states, additionally a squeezed gain $\hat{\sigma}^\dagger$ is present.


\subsection{Noise and moments taken \label{subsec:noiseandmoments}}

\begin{figure}
    \centering
    \includegraphics[width=\linewidth]{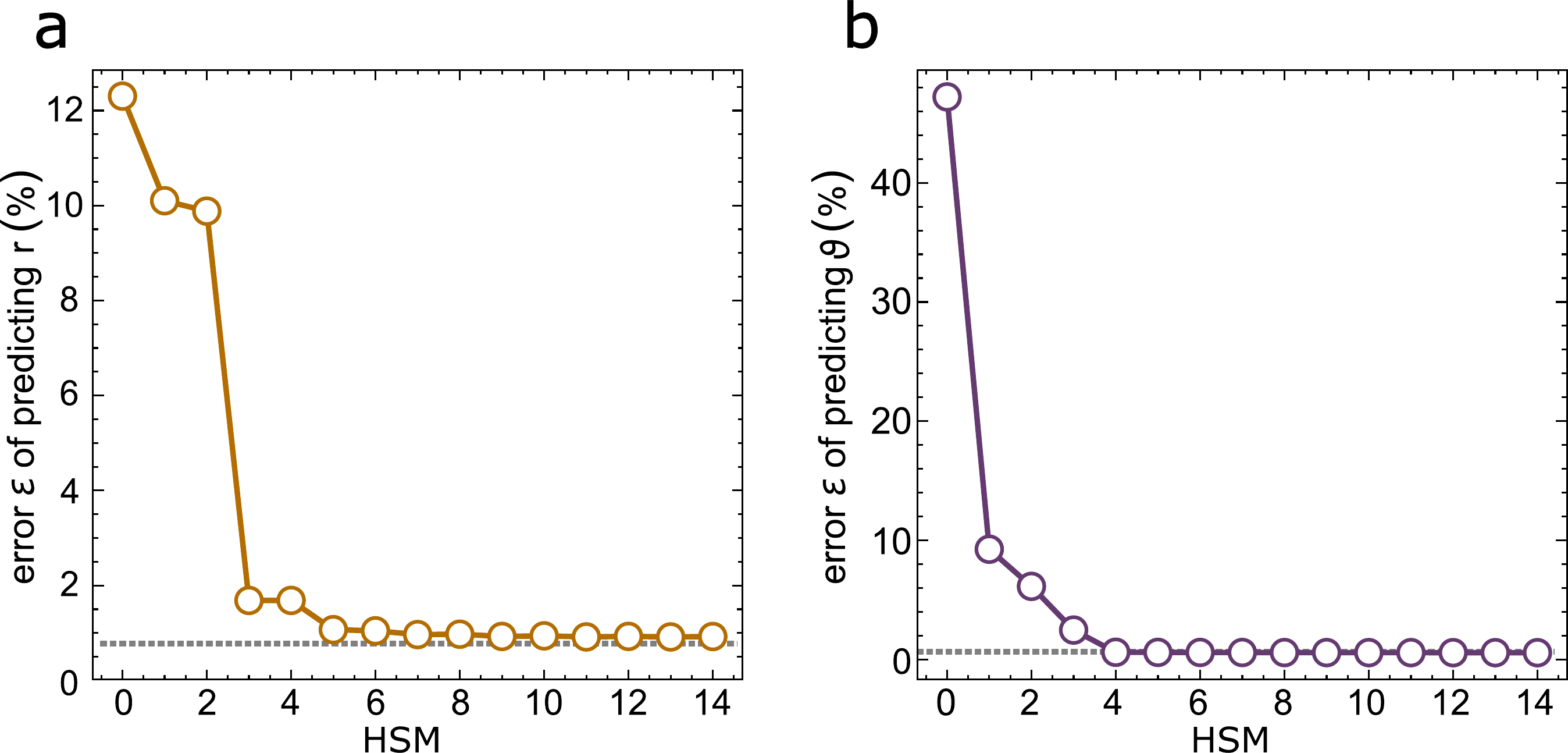}
    \caption{Error of predicting $r$ (panel \textbf{a}) and $\theta$ (panel \textbf{b}) as a function of the highest spectral moment (HSM) considered for regression. Within moments up to order $m\leq 4$, we observe a reasonable convergence to the result of the full spectrum (gray, dashed). Ridge parameter $10^{-14}$ was used to avoid large deviations for the highest number of moments or the full spectrum. }
    \label{fig:Momentscandprediction}
\end{figure}

We aim to have a better understanding of the selection of the output nodes, focusing on the tasks of predicting $r$ or $\theta$ as in \ref{subsec:squeezspectra}.
In Fig \ref{fig:Momentscandprediction} we compare the performance of the reservoir trained by its $m+1$ first moments versus one trained with the whole spectra. Taking the full spectra is generally able to produce the lowest value of $\varepsilon$, but at the cost of an extensive number of variables. As we also see, the first few moments (e.g. $m\in\{0,\ldots,4\}$) approximate this result effficiently.

\begin{figure}
    \centering
    \includegraphics[width=\linewidth]{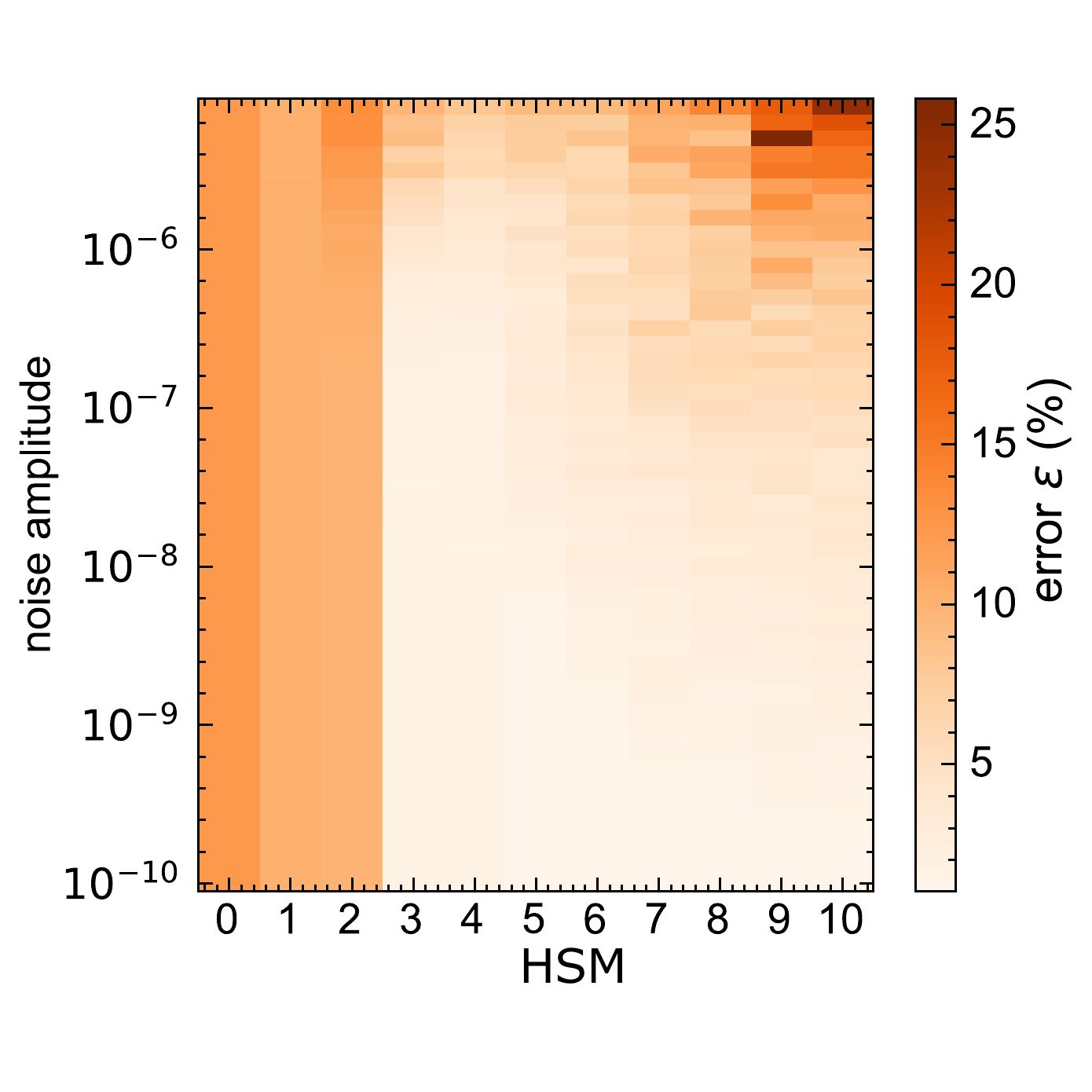}
    \caption{Dependence of the prediction error $\varepsilon$ on the presence of external noise in the spectrum, during estimation of $r$. Whereas considering moments with $m^\mathrm{max}\leq 2$ always leads to a significant value of $\varepsilon$, moderate value of $m^\mathrm{max}$ give an optimal result when the noise becomes significant.}
    \label{fig:noisedependence}
\end{figure}

In Fig \ref{fig:noisedependence}, the vulnerability to noise is studied. While we retrieve the finding that the performance improves initially with more moments, taking too many moments leads to a detoriation in presence of substantial noise, which can be attributed to overfitting

Together, these observations corroborate the motivation to focus on working with moments $M_0,\ldots,M_4$ in the majority of cases.

\subsection{OPO and polariton model\label{subsec:OPOmodel}}

We now move on to the complete description of squeezed state generation by an OPO with cascaded coupling into the microcavity. Solving this combined system exactly in a truncated Fock space would be very memory demanding. Instead, we will work in the Positive P method \cite{Deuar2021,gardiner2004quantum}, which turns out to have the additional benefit that the dynamics of source and microcavity decouple. This method has also been shown to be suitable for the description of polariton systems with many modes for the context of reservoir computing \cite{Swierczewski25}.

In the positive P framework, the model of the OPO (Eq. \ref{eq:OPOham}) gives rise to the equations

\begin{align}\label{eq:OPOposPs}
 \partial_t\phi&=-\frac{\gamma_s}{2}\phi+G\tilde{\phi}^*+\sqrt{G}\xi_s  \ao{,}\\
  \partial_t\tilde{\phi}^*&=-\frac{\gamma_s} {2}\tilde{\phi}^*+G^*\phi+\sqrt{G^*}\tilde{\xi_s}, 
\end{align}
where $\xi,\tilde{\xi}$ are real-valued stochastic (Wiener) processes. Consistent with the input-output formalism \cite{Walls_Milburn_2008}, the output fields of the OPO are $\phi_\text{out}=\sqrt{\gamma_s\epsilon_s}\phi$ and $\tilde{\phi}^*_\text{out}=\sqrt{\gamma_s\epsilon_s}\tilde{\phi}^*$. To create a displaced (bright) squeezed state, another coherent beam $\alpha$ at the same frequency as the central frequency of the squeezed state can  be added in a beamsplitter. The input to the polariton system is then $\psi_\text{IN}=\phi_\text{OUT}+\alpha$ and $\tilde{\psi}^*_\text{IN}=\tilde{\phi}^*_\text{OUT}+\alpha^*$.

The Positive P equations of the polariton cavity take the form \cite{Swierczewski25}
\begin{align}\label{eq:PolposPs}
\partial_t \psi=&\left(-iU \tilde{\psi}^*\psi-\frac{\gamma_c}{2}+\sqrt{-iU}\xi_p\right)\psi\nonumber\\ &+\sqrt{\epsilon_p\gamma_c}\psi_\text{IN}(t)\\
\partial_t \tilde{\psi}^*=&\left(iU \tilde{\psi}^*\psi-\frac{\gamma_c}{2}+\sqrt{iU}\tilde{\xi_p}\right)\tilde{\psi}^*\nonumber \\ &+\sqrt{\epsilon_p\gamma_c}\tilde{\psi}^*_\text{IN}(t),
\end{align}

where an overall, unimportant phase $\pi$ was absorbed in the last terms. The parameters $0\leq\epsilon_{s,p}\leq 1$ characterize how much the OPO and cavity couple to each other, as a fraction to their total range of interaction with the environment. 

 A seperation of timescales where $\gamma_c \gg \gamma_s$ allows us to make a quasi-static approximation. The dynamics of the source \eqref{eq:OPOposPs} can be solved first. For each corresponding value of $\psi_\text{IN}$, the polariton dynamics \eqref{eq:PolposPs} can then be studied until the steady state regime. 

This model could be similarly described with other phase-space methods such as the truncated Wigner approximation, which is likely adequate despite formally inexact. However, only the P distributions readily yield the time-and-normally-ordered correlation functions as needed for the spectra ( Eq. \eqref{eq:specintegral}) \cite{Walls_Milburn_2008,Deuar2021multitime}.

Parameters used correspond to those achieved in Ref \cite{Burks09} ($\epsilon_s=0.7, \gamma_s=10~ ms^{-1}, \abs{G}=4.5 ms^{-1}$), while $\epsilon_p=0.45$ was estimated.
For the numerical results presented in Fig \ref{fig:OPO}, eqs. \eqref{eq:OPOposPs} for the OPO were, starting from $\phi=\tilde{\phi}^*=$, freely evolved for $20 \gamma_s^{-1}$, and from there, the resulting $(\phi_\mathrm{out},\tilde{\phi}^*_\mathrm{out})$ were saved every $\gamma_s^{-1}$ for a total OPO evolution of $1000\gamma_s^{-1}$. From each such $(\phi_\mathrm{out},\tilde{\phi}^*_\mathrm{out})$,   $(\psi_\mathrm{in},\tilde{\psi}^*_\mathrm{in})$ were initialized. The microcavity evolution \eqref{eq:PolposPs} for each of them runs for an initial $100 \gamma_c^{-1}$ to relax from the initial vacuum state, and then a subsequent $200 \gamma_c^{-1}$ during which the output is stored every $0.5\gamma_c^{-1}$. This brings the total equivalent evolution time for a single Positive P trajectory to $2\times 10^5\gamma_c^{-1}\approx 3 \mu s$. The total evolution was paralellized over $10^3$ positive P trajectories. A single experimental run, does not, even under idealised measurements where the quantum state remains pure, correspond to only a single phase space trajectory for its evolution \cite{gardiner2004quantum} however. This means that the total integrated evolution time $3~\mu s<T< 3~ms$.


\subsection{Spectral data renormalization for multivariate prediction\label{subsec:multimet}}

To prepare the spectral data for optimal learning, the spectrum is initially \textit{binned} (meaning the data points of the spectrum are divided into smaller intervals, with the average value over this interval constituting to a new datapoint in the binned spectrum). Such treatment decreases the number of datapoints while preserving the shape of the spectrum. It also decreases (due to averaging within the bins) the impact of any random noise, which may be present in the spectral data. Second, because the amplitude of the spectra spans several orders of magnitude, renormalizing it makes it easier to process for a neural network, as the optimalisation algorithm and neuron activation function performs better for a normalised input signal. A randomly chosen spectrum from the dataset is chosen as a reference $S_{ref}(\omega)$, and every other spectrum (both in the training as in the testing dataset) is renormalized by taking a difference of it and the reference and dividing by the mean value of the reference spectrum. $S_{i}(\omega) \rightarrow (S_{i}(\omega) - S_{ref}(\omega))/\left\langle S_{ref} (\omega)\right\rangle$ (see Fig. \ref{fig:enhanced}\textbf{c}).

From these renormalized spectral data, we construct the training $\mathbf{D}_{\text{train}} = \{(\mathbf{x}_i, \mathbf{y}_i)\}_{i=1}^{N{\text{train}}}$ and testing datasets $\mathbf{D}_{\text{test}} = \{(\mathbf{x}_i, \mathbf{y}_i)\}_{i=1}^{N{\text{test}}}$, containing $N_{\text{train}} = 700$ and $N_{\text{test}} = 100$ samples. Each feature vector $\mathbf{x}_i$ encodes the pre-processed spectrum corresponding to a given squeezed state, while each target value $\mathbf{y}_i$ consists of the corresponding squeezing parameter $r$ and phase $\theta$.

The procedures described above define the complete workflow of our study, from the theoretical modeling of the squeezed environment to the numerical treatment of the OPO-driven cavity spectra and their post-processing for machine learning. Having established this methodological basis, we conclude by summarizing the main insights and implications of our results for quantum reservoir computing.

\section{Conclusion}

Reservoir computing constitutes a promising avenue for the characterization of quantum states such as squeezed states that is gaining increasing attention \cite{Swierczewski25,Regemortel2025,Ko2025}. In this work, we have shown that this could be feasible with a reservoir as simple as a single nonlinear microcavity, from which the emission spectrum is taken as the output nodes. Rather than idealised single-mode Gaussian states, we considered continuous sources. Good prediction accuracy was achieved for the different parameters characterizing the initial state. Nonlinearity in the reservoir was confirmed to play a significant role, with the realistic values of the nonlinearity in the theoretically
optimal range. Learning of the squeezing parameter has shown to generalise across other parameter values. Although some variables are more easily extracted from a multivariate dataset than others with a simple regression on the output spectrum, we have shown that an additional classical feed-forward neural network leads to a very good performance on all variables. 
Finally, we have shown that this potential for reservoir computing is also present for an inital squeezed state state from an explicitly modeled realistic OPO, even if the latter has very different specifications from the cavity reservoir. While we have shown it to be not strictly necessary, future improvements may be possible through additional features such as time lenses that attune the dynamics of source and cavity reservoir \cite{Foster2009,Joshi22}. We have all reasons to believe that this approach will work well on other quantum states, such as single-photon states, which are number squeezed, as well.

These results highlight the potential of reservoir computing with simple setups as well as a role that quantum polaritonics \cite{Sanvitto2016} can assume.

\begin{acknowledgement}

This work was financed by the European Union EIC Pathfinder Challenges project ``Quantum Optical Networks based on Exciton-polaritons'' (Q-ONE, Id: 101115575). The authors gratefully thank the other consortium partners for the many stimulating discussions. AO acknowledges support from the National Science Center, Poland, project No. 2024/52/C/ST3/00324.

\end{acknowledgement}




\bibliography{bibliography}

\end{document}